\documentclass[aps, prd, reprint, twocolumn, superscriptaddress]{revtex4-1}
\usepackage{amsmath}
\usepackage{amssymb}
\usepackage{graphicx}
\usepackage{color}
\usepackage{times}
\usepackage{euscript}
\usepackage{amsthm}
\usepackage{amsfonts}
\usepackage[english]{babel}
\usepackage{epstopdf}
\usepackage{multirow,makecell,array}
\usepackage{mathtools}
\usepackage{float}
\usepackage[dvipsnames]{xcolor}

\begin{document}

\title{Vacuum birefringence and dichroism in a strong plane-wave background}

\author{I.~A.~Aleksandrov}
\affiliation{Department of Physics, Saint Petersburg State University, Universitetskaya Naberezhnaya 7/9, Saint Petersburg 199034, Russia}
\affiliation{Ioffe Institute, Politekhnicheskaya street 26, Saint Petersburg 194021, Russia}
\author{V.~M.~Shabaev}
\affiliation{Department of Physics, Saint Petersburg State University, Universitetskaya Naberezhnaya 7/9, Saint Petersburg 199034, Russia}
\affiliation{National Research Centre ``Kurchatov Institute'' B.P. Konstantinov Petersburg Nuclear Physics Institute, Gatchina, Leningrad district 188300, Russia}

\begin{abstract}
In the present study, we consider the effects of vacuum birefringence and dichroism in strong electromagnetic fields. According to quantum electrodynamics, the vacuum state exhibits different refractive properties depending on the probe photon polarization and one also obtains different probabilities of the photon decay via production of electron-positron pairs. Here we investigate these two phenomena by means of several different approaches to computing the polarization operator. The external field is assumed to be a linearly polarized plane electromagnetic wave of arbitrary amplitude and frequency. Varying the probe-photon energy and the field parameters, we thoroughly examine the validity of the locally-constant field approximation (LCFA) and techniques involving perturbative expansions in terms of the external-field amplitude. Within the latter approach, we develop a numerical method based on a direct evaluation of the weak-field Feynman diagrams, which can be employed for investigating more complex external backgrounds. It is demonstrated that the polarization operator depends on two parameters: classical nonlinearity parameter $\xi$ and the product $\eta = \omega q_0 / m^2$ of the laser field frequency $\omega$ and the photon energy $q_0$ ($m$ is the electron mass). The domains of validity of the approximate techniques in the $\xi \eta$ plane are explicitly identified.
\end{abstract}

\maketitle

\section{Introduction}\label{sec:intro}

According to quantum electrodynamics (QED), the physical vacuum state contains quantum fluctuations of the electromagnetic and electron-positron fields, which can be viewed as spontaneous creation and annihilation of electron-positron pairs interacting with each other via virtual photons. Although these virtual particles are not observable themselves, their existence can manifest itself while interacting with external fields and real particles giving rise to a number of remarkable nonlinear phenomena such as light-by-light scattering~\cite{euler_kockel, heisenberg_euler, weisskopf, karplus_pr_1950_1951}, Sauter-Schwinger effect~\cite{heisenberg_euler, sauter_1931, schwinger_1951}, and so on (for review, see, e.g., Refs.~\cite{dipiazza_rmp_2012, xie_review_2017, fedotov_review}). In this investigation, we consider propagation of a probe photon in vacuum in the presence of a strong external background. The latter polarizes the physical vacuum, so the probe photon effectively interacts with a nonlinear medium, which leads to the phenomena of vacuum birefringence and dichroism~\cite{Toll:1952, Baier, Baier2, baier_1976, becker_1975} which are in the focus of the present study (we note that the nontrivial properties of the vacuum state in the presence of real photons give also rise to recently discussed stimulated photon emission~\cite{aleksandrov_antonino_2022}).

Observing these processes in the laboratory represents currently an intriguing and challenging task. There are mainly two different approaches to probing vacuum birefringence. First, one can rely on unprecedented accuracy of experimental measurements in the optical domain, i.e., in the regime of relatively low probe-photon energies (see, e.g., Refs.~\cite{di_piazza_prl_2006, heinzl_2006, dinu_prd_2014, karbstein_zepf_2015, felix_mosman_prd_2020, felix_prl_2022}). From the theoretical viewpoint, this domain allows one to employ local approximations, i.e., to treat the external (laser) field as a locally constant background. The corresponding locally-constant field approximation (LCFA) has basically two different implementations based either on employing the exact expressions for the Heisenberg-Euler effective Lagrangian~\cite{felix_rashid_2015} or on using the local values of the polarization operator derived in constant crossed fields~\cite{meuren_2013, bragin_2017}. The second approach to vacuum birefringence involves high-energy probe photons~\cite{king_elkina_2016, nakamiya_2017, bragin_2017}. The advantage of this technique appears due to large probabilities of the corresponding quantum processes resulting in large values of the experimental signal. On the other hand, it is significantly more difficult to perform measurements in the high-energy domain as, e.g., the Heisenberg-Euler approximation is only valid in the low-energy domain. To properly assess the feasibility of the corresponding scenarios, one has to obtain accurate and reliable theoretical predictions.

In order to avoid approximate local treatment of the external electromagnetic field, one can model it with a plane-wave background allowing one to deduce explicit analytical expressions for the polarization tensor~\cite{baier_1976, becker_1975, meuren_2013}. On the other hand, this simplified setup may not properly reflect the properties of real experimental conditions.

In the present study, we have two primary aims. First, we will thoroughly examine the plane-wave scenario by means of analytical nonperturbative expressions derived in Refs.~\cite{baier_1976, becker_1975, meuren_2013}. We will compute the polarization tensor in a wide range of physical parameters governing the process under consideration: laser-field amplitude, laser frequency, and probe-photon energy. Expanding the nonperturbative result in powers of the external-field amplitude, we will assess the accuracy of the calculations based on perturbation theory (PT). Besides, we will quantitatively analyze the validity of the LCFA in the two forms described above. Second, the polarization tensor will be directly evaluated via the corresponding Feynman diagrams. This approach is very important since it can allow one to consider other field configurations, which differ from a simple plane-wave scenario. In what follows, we will benchmark our direct computational procedures and also provide an additional insight into the analytical properties of the integrands involved in the Feynman diagrams. For instance, it will be demonstrated that the overlap between the branch cuts that appears for sufficiently high photon energies is closely related to the decay of the probe photon via production of electron-positron pairs. We also mention that $e^+e^-$ pairs can be produced directly by a classical strong field, i.e., via the Sauter-Schwinger mechanism. The validity of the LCFA in this context was recently examined in Refs.~\cite{aleksandrov_prd_2019_1, sevostyanov_prd_2021, aleksandrov_symmetry, aleksandrov_sevostyanov_2022}.

The paper has the following structure. In Sec.~\ref{sec:setup} we describe the setup under consideration involving a probe photon and external plane-wave background. In Sec.~\ref{sec:np_analytic} we present nonperturbative expressions which we employ in our numerical computations. In Sec.~\ref{sec:pt} we calculate the leading-order contribution with respect to the external-field amplitude. Section~\ref{sec:lcfa} is devoted to the description of the two possible implementations of the LCFA. In Sec.~\ref{sec:diagrams} we discuss how one can directly evaluate the leading-order Feynman diagrams. Section~\ref{sec:results} contains our numerical results obtained by means of the various techniques. Finally, we conclude in Sec.~\ref{sec:conclusions}.

Throughout the text, we employ the units $\hbar = c = 1$, $\alpha = e^2/(4\pi) \approx 1/137$.

\section{Setup and notation}\label{sec:setup}

We assume that the external plane wave is polarized along the $x$ axis and propagates in the $z$ direction, i.e., it depends on $\varphi = \omega n^\mu x_\mu = \omega (t-z)$, where $\omega$ is the laser frequency. The null vector $n$ obeys $n_0 = 1$, $n^2 = 0$. The corresponding vector potential has the following form:
\begin{eqnarray}
\boldsymbol{A} (x) &=& \mathcal{A} (\omega(t-z)) \boldsymbol{e}_x, \label{eq:A_gen} \\
\mathcal{A} (\varphi) &=& \frac{E_0}{\omega} \sin \varphi,
\label{eq:A_pot}
\end{eqnarray}
where $E_0$ is the field strength amplitude. We also introduce a dimensionless parameter $\xi = |eE_0|/(m\omega)$. The initial photon momentum $\boldsymbol{q}$ points in the opposite direction to $\boldsymbol{n} = \boldsymbol{e}_z$, $\boldsymbol{q} = -q_0 \boldsymbol{e}_z$. Accordingly, the initial 4-momentum of the photon is $q^\mu = q_0 (1, 0, 0, -1)^\text{t}$. The final momentum will be denoted by $k^\mu$. In what follows, we will also employ the light-cone components which for arbitrary 4-vector $v^\mu$ read
\begin{eqnarray}
v_+ &=& \frac{v_0 + \boldsymbol{n}\boldsymbol{v}}{2},\\
v_- &=& v_0 - \boldsymbol{n}\boldsymbol{v},\\
\boldsymbol{v}_\perp &=& \boldsymbol{v} - (\boldsymbol{n} \boldsymbol{v}) \boldsymbol{n}. 
\end{eqnarray}
The scalar product of two vectors can be evaluated via
\begin{equation}
vw \equiv v^\mu w_\mu = v_+w_- + v_- w_+ - \boldsymbol{v}_\perp \boldsymbol{w}_\perp.
\label{eq:light-cone_SP}
\end{equation}
For instance, $n_+ = 1$, $n_- = 0$, $\boldsymbol{n}_\perp = 0$, and $\varphi = \omega x_-$.

\begin{figure}[t]
  \center{\includegraphics[width=0.6\linewidth]{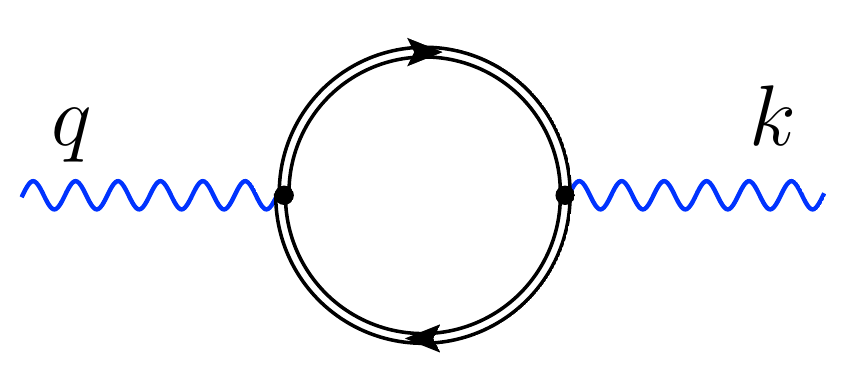}}
  \caption{Feynman diagram describing the leading-order contribution to the photon polarization operator. The amplitude of the process is proportional to the fine-structure constant $\alpha$ and exactly takes into account the interaction with the classical external background (double lines represent the dressed Green's functions).}
  \label{fig:diag_gen}
\end{figure}

The amplitude $\mathcal{S}(q, k)$ of the process described by the diagram in Fig.~\ref{fig:diag_gen} involves two photon wavefunctions defined as
\begin{equation}
f^\mu_{q} (x) = \frac{1}{\sqrt{2q_0}} \mathrm{e}^{-iqx} \varepsilon^\mu (q),
\label{eq:photon_wf}
\end{equation}
where $\varepsilon^\mu (q)$ is the polarization 4-vector. The amplitude can be represented in the form
\begin{equation}
\mathcal{S}(q, k) = \frac{1}{\sqrt{4q_0k_0}} \varepsilon_\mu (q) i \big [\Pi_0^{\mu \nu} (q, k) + \Pi^{\mu \nu} (q, k) \big ] \varepsilon^*_\nu (k).
\end{equation}
Here $\Pi_0^{\mu \nu} (q, k)$ denotes the zero-field contribution to the polarization operator, which corresponds to the diagram with the free-electron Green's functions describing vacuum polarization in the absence of external fields. This contribution diverges and requires a usual renormalization procedure. Since this term does not affect the processes of vacuum birefringence and dichroism, our task is to compute the field-dependent part $\Pi^{\mu \nu} (q, k)$, which is finite.

In what follows, we will evaluate $\Pi^{\mu \nu} (q, k)$ by means of several different techniques mentioned above. As will be seen below, the polarization operator involving $\xi$, $\omega$, and $q_0$ depends, in fact, only on $\xi$ and the product $\omega q_0$. We will consider $\xi$ and $\eta \equiv \omega q_0/m^2$ as two independent dimensionless parameters governing the processes of vacuum birefringence and dichroism. We will also introduce the so-called quantum nonlinearity parameter $\chi = 2\xi \eta$ which will be considered as a derived quantity $\chi (\xi, \eta)$.

\section{Nonperturbative analytical formulas}\label{sec:np_analytic}

In the case of a plane-wave external background, it is possible to compute the polarization tensor analytically. In Ref.~\cite{baier_1976} it was done by means of the operator approach. In Ref.~\cite{becker_1975} the calculations were performed in the case of a monochromatic plane wave. Recently, in Ref.~\cite{meuren_2013} the results of Refs.~\cite{baier_1976, becker_1975} were confirmed by direct computations of the Feynman diagram in Fig.~\ref{fig:diag_gen} with the aid of the exact Green’s functions, which can be constructed from the Volkov solutions.

Here we will first employ the general expressions presented in Refs.~\cite{baier_1976, becker_1975, meuren_2013}. Due to the symmetry of the external plane-wave field, it can only change the $q_+$ component of the photon momentum, so the amplitude corresponding to the Feynman diagram in Fig.~\ref{fig:diag_gen} contains $\delta (k_- - q_-) \delta(\boldsymbol{k}_\perp - \boldsymbol{q}_\perp)$. It turns out that the cumbersome expressions for the amplitude derived in Refs.~\cite{baier_1976, becker_1975, meuren_2013} become relatively simple in the particular case of a circularly polarized plane-wave background. Due to the helicity conservation, the momentum component $q_+$ can change only by $\pm 2\omega$ or remain the same. It is not the case if the external field has a linear polarization since such a plane wave does not possess a well-defined helicity quantum number. Accordingly, the $q_+$ momentum component of the photon may change by an arbitrary integer number of $\omega$. The general expression for the setup described above has the following form:
\begin{widetext}
\begin{equation}
\Pi^{\mu \nu} (q, k) = -\frac{4\pi^2\alpha}{\omega} \delta (k_- - q_-) \delta(\boldsymbol{k}_\perp - \boldsymbol{q}_\perp) \int \limits_{-1}^{1} dv \int \limits_{0}^{\infty} \frac{d\tau}{\tau} \int \limits_{-\infty}^{\infty} d\varphi \, \mathrm{e}^{i\Phi} \begin{pmatrix} c & 0 & 0 & 0 \\ 0 & b + \Delta b & 0 & 0 \\ 0 & 0 & b & 0 \\ 0 & 0 & 0 & c \end{pmatrix},
\label{eq:np_gen} 
\end{equation}
where
\begin{eqnarray}
b &=& \Big ( \frac{i}{\tau} + \frac{1}{2}kq \Big ) (1-\mathrm{e}^{i\tau \beta}) + \frac{2m^2 \tau \xi^2}{\mu} \, \mathrm{e}^{i\tau \beta} \sin^2 (\mu \omega q_0) \cos^2 \varphi, \label{eq:gen_b} \\
\Delta b &=& 2m^2 \xi^2 \Big [ \mathrm{sinc}^2 (\mu \omega q_0) \sin^2 \varphi - 2 \, \mathrm{sinc} (2 \mu \omega q_0) \sin^2 \varphi - \sin^2 (\mu \omega q_0) + \sin^2 \varphi \Big ] \mathrm{e}^{i\tau \beta}, \label{eq:gen_delta_b} \\
c &=& \frac{k_0q_0\mu}{\tau} \, (1-\mathrm{e}^{i\tau \beta}), \label{eq:gen_c}\\
\mu &=& \frac{1}{2} \tau (1-v^2), \label{eq:gen_mu}\\
\Phi &=& \frac{k_+ - q_+}{\omega} \varphi + \frac{1}{2} \mu kq - m^2 \tau, \label{eq:gen_phi}\\
\beta &=& m^2 \xi^2 \bigg [ \mathrm{sinc}^2 (\mu \omega q_0) \sin^2 \varphi - \frac{1}{2} + \frac{1}{2} \, \mathrm{sinc} (2\mu \omega q_0) \cos 2\varphi \bigg ]. \label{eq:gen_beta}\\
\end{eqnarray}
In what follows, we will be interested only in the elastic process, where $k_+ = q_+$ as the other channels are significantly suppressed (actually, they rather represent reactions involving photon merging or splitting than the phenomenon of birefringence). To extract the particular process of elastic scattering, one has to isolate the zeroth-order Fourier harmonics with respect to $\varphi$ dependence in the functions $b$, $\Delta b$, and $c$, so the integration of $\mathrm{exp} (i\Phi)$ yields the necessary delta-function. This can be straightforwardly attained with the aid of the Jacobi-Anger identity. The result reads
\begin{equation}
\Pi_\text{elastic}^{\mu \nu} (q, k) = - (2\pi)^3 \alpha \delta (k - q) \int \limits_{-1}^{1} dv \int \limits_{0}^{\infty} \frac{d\tau}{\tau} \, \mathrm{e}^{-im^2 \tau} \begin{pmatrix} \tilde{c} & 0 & 0 & 0 \\ 0 & \tilde{b} + \Delta \tilde{b} & 0 & 0 \\ 0 & 0 & \tilde{b} & 0 \\ 0 & 0 & 0 & \tilde{c} \end{pmatrix},
\label{eq:np_elastic} 
\end{equation}
where
\begin{eqnarray}
\tilde{b} &=& \frac{i}{\tau} [ 1 - \Xi J_0(A) ] + \frac{m^2 \tau \xi^2}{\mu} \sin^2 (\mu \omega q_0) \Xi [J_0 (A) + iJ_1 (A)], \label{eq:el_b} \\
\Delta \tilde{b} &=& m^2 \xi^2 \Xi \big \{ -2 \sin^2 (\mu \omega q_0) J_0 (A) + [\mathrm{sinc}^2 (\mu \omega q_0) - 2 \, \mathrm{sinc} (2 \mu \omega q_0) +1 ] [J_0 (A) - iJ_1 (A)] \big \}, \label{eq:el_delta_b} \\
\tilde{c} &=& \frac{q^2_0\mu}{\tau} \, [1-\Xi J_0 (A)], \label{eq:el_c}\\
\Xi &=& \mathrm{exp} \bigg \{ \frac{i}{2} m^2 \tau \xi^2 [\mathrm{sinc}^2 (\mu \omega q_0) - 1]\bigg \}, \label{eq:el_xi}\\
A &=& \frac{1}{2} m^2 \tau \xi^2 [\mathrm{sinc} (2 \mu \omega q_0) - \mathrm{sinc}^2 (\mu \omega q_0)]. \label{eq:el_A}
\end{eqnarray}
\end{widetext}
Here $J_n$ are the Bessel functions of the first kind. We will assume hereinafter $k^\mu = q^\mu$. We also note that the elements $\Pi^{00}$ and $\Pi^{33}$ are equal, which preserves the gauge invariance and the Ward-Takahashi identity~\cite{blp}. These components will not be evaluated in our study as they do not affect the phenomena under consideration.

The birefringent and dichroic properties of the vacuum in the presence of strong fields manifest themselves in the difference between $\Pi^{11}$ and $\Pi^{22}$ elements: photon polarizations along the $x$ and $y$ axes correspond to different refractive and absorption indexes. In what follows, we will compute these elements. As was stated above, these quantities involve the three parameters $\xi$, $\omega$, and $q_0$, but they depend, in fact, on $\xi$ and $\eta = \omega q_0 / m^2$ as becomes evident from Eqs.~\eqref{eq:np_elastic}--\eqref{eq:el_A}.

\begin{figure*}[t]
  \center{\includegraphics[width=0.85\linewidth]{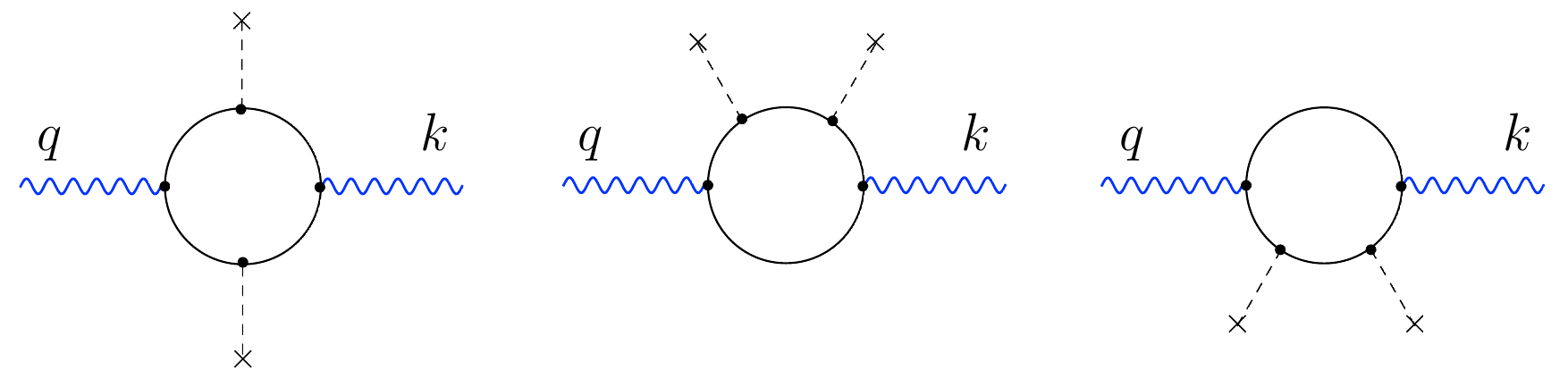}}
  \caption{Feynman diagrams corresponding to the leading-order contribution within the PT expansion in terms of the external field (the amplitudes are proportional to $\xi^2$). The interaction with the classical external field is denoted by the cross. Depending on the energy-momentum transfer at the cross vertices, the process is either elastic (2-to-2 process) or corresponds to $k \neq q$.}
  \label{fig:diag_three}
\end{figure*}

\section{Perturbation theory}\label{sec:pt}

Here we will consider the leading-order term of Eq.~\eqref{eq:np_elastic} with respect to the small-$\xi$ expansion. This contribution is proportional to $\xi^2$ and corresponds to the three Feynman diagrams displayed in Fig.~\ref{fig:diag_three}. Expanding the function $\Xi$ and the Bessel functions in Taylor series, one obtains
\begin{eqnarray}
\tilde{b}_\text{LO} &=& m^2 \xi^2 \bigg \{ \frac{1}{2} [\mathrm{sinc}^2 (\mu \omega q_0) - 1] + \frac{\tau}{\mu} \, \sin^2 (\mu \omega q_0) \bigg \}, \label{eq:el_b_lo} \\
\Delta \tilde{b}_\text{LO} &=& m^2 \xi^2 [ -2 \sin^2 (\mu \omega q_0) + \mathrm{sinc}^2 (\mu \omega q_0) \nonumber \\
{}&-& 2 \, \mathrm{sinc} (2 \mu \omega q_0) +1 ], \label{eq:el_delta_b_lo} \\
\tilde{c}_\text{LO} &=& -\frac{i}{2} q_0^2\mu m^2 \xi^2 [\mathrm{sinc}^2 (\mu \omega q_0) - 1]. \label{eq:el_c_lo}
\end{eqnarray}
Here ``LO'' stands for ``low order''. It turns out that one can replace $\mu$ with Eq.~\eqref{eq:gen_mu} and perform the $\tau$ integration analytically. Let us first introduce the following general representation:
\begin{equation}
\Pi_\text{elastic}^{\mu \nu} (q, k) = - (2\pi)^3 \alpha \delta (k - q) m^2 \xi^2 M^{\mu \nu}. \label{eq:M_gen}
\end{equation}
Within PT we find
\begin{eqnarray}
M^{11}_\text{LO} &=& \int \limits_{-1}^{1} dv \bigg [ \frac{2v^2}{1-v^2} \, I_1 (v) + \frac{1}{2} \, I_2 (v) + I_3 (v) \bigg ],  \label{eq:M11} \\
M^{22}_\text{LO} &=& \int \limits_{-1}^{1} dv \bigg [ \frac{2}{1-v^2} \, I_1 (v) + \frac{1}{2} \, I_2 (v) \bigg ],  \label{eq:M22}
\end{eqnarray}
where
\begin{widetext}
\begin{eqnarray}
I_1(v) &=& \int \limits_0^{\infty} \frac{dt}{t} \, \sin^2 (\gamma t) \mathrm{e}^{-it} = \frac{1}{4} \ln \big | 1 - 4\gamma^2 \big | - \frac{i\pi}{4} \theta (\gamma - 1/2), \label{eq:I1}\\
I_2(v) &=& \int \limits_0^{\infty} \frac{dt}{t} \, [\mathrm{sinc}^2 (\gamma t) -1] \mathrm{e}^{-it} = \frac{3}{2} - \frac{1}{2} \bigg ( 1 + \frac{1}{4\gamma^2} \bigg ) \ln \big | 1 - 4\gamma^2 \big | - \frac{1}{2\gamma} \, \ln \bigg | \frac{1 + 2\gamma}{1 - 2 \gamma} \bigg | \nonumber \\
{}&+& \frac{i\pi}{2} \bigg ( 1 - \frac{1}{\gamma} + \frac{1}{4\gamma^2} \bigg ) \theta (\gamma - 1/2), \label{eq:I2}\\
I_3(v) &=& \int \limits_0^{\infty} \frac{dt}{t} \, [1 + \mathrm{sinc}^2 (\gamma t) - 2 \, \mathrm{sinc} (2\gamma t)] \mathrm{e}^{-it} = -\frac{1}{2} + \frac{1}{2} \bigg ( 1 - \frac{1}{4\gamma^2} \bigg ) \Big [ \ln \big | 1 - 4\gamma^2 \big | - i\pi \theta (\gamma - 1/2) \Big ], \label{eq:I3}\\
\gamma &=& \gamma (v) = \frac{\omega q_0}{2m^2} (1-v^2) = \frac{1}{2} \eta (1-v^2).
\end{eqnarray}
The expressions \eqref{eq:M11} and \eqref{eq:M22} depend only on $\eta = \omega q_0/m^2$, while the nonperturbative values of $M^{\mu\nu}$ [see Eq.~\eqref{eq:M_gen}] also involve $\xi$. Below we will compare the leading-order terms with the nonperturbative results. Let us now present the low- and high-energy asymptotic expressions for $M^{11}_\text{LO}$ and $M^{22}_\text{LO}$. In the low-energy case $\varepsilon \equiv 2 \eta = 2\omega q_0/m^2 \ll 1$,
\begin{eqnarray}
M^{11}_\text{LO} &=& -\frac{4}{45}\varepsilon^2 -\frac{17}{3150}\varepsilon^4 + \mathcal{O} (\varepsilon^6), \label{eq:M11_asym_low_en}\\
M^{22}_\text{LO} &=&  -\frac{7}{45}\varepsilon^2 -\frac{131}{9450}\varepsilon^4 + \mathcal{O} (\varepsilon^6). \label{eq:M22_asym_low_en}
\end{eqnarray}
In the high-energy limit, we obtain [$\varepsilon \equiv 1/(2\eta) = m^2/ (2 \omega q_0) \ll 1$]
\begin{eqnarray}
M^{11}_\text{LO} &=& \frac{1}{2}\ln^2 \varepsilon + \bigg ( 1 - \ln 2 + \frac{i\pi}{2} \bigg ) \ln \varepsilon + \bigg [ \frac{5}{2} - \ln 2 + \frac{1}{2} \ln^2 2 - \frac{\pi^2}{4} + \frac{i\pi}{2} (1 - \ln 2) \bigg ] \nonumber \\
{}&+& i\pi \varepsilon \ln \varepsilon + \bigg [-\frac{\pi^2}{2} + \frac{i\pi}{2} (3 - 2 \ln 2) \bigg ] \varepsilon + \mathcal{O} (\varepsilon^2 \ln^2 \varepsilon), \label{eq:M11_asym}\\
M^{22}_\text{LO} &=&  \frac{1}{2}\ln^2 \varepsilon + \bigg ( 1 - \ln 2 + \frac{i\pi}{2} \bigg ) \ln \varepsilon + \bigg [ \frac{7}{2} - \ln 2 + \frac{1}{2} \ln^2 2 - \frac{\pi^2}{4} + \frac{i\pi}{2} (1 - \ln 2) \bigg ] \nonumber \\
{}&+& i\pi \varepsilon \ln \varepsilon + \bigg [-\frac{\pi^2}{2} + \frac{i\pi}{2} (1 - 2 \ln 2) \bigg ] \varepsilon + \mathcal{O} (\varepsilon^2 \ln^2 \varepsilon). \label{eq:M22_asym}
\end{eqnarray}
While the low-energy result~\eqref{eq:M11_asym_low_en}, \eqref{eq:M22_asym_low_en} is real, the expressions~\eqref{eq:M11_asym} and \eqref{eq:M22_asym} possess imaginary parts, which describe the process of photon decay. The imaginary part of the difference $\delta M_\text{LO} \equiv M^{11}_\text{LO} - M^{22}_\text{LO} \approx -1 + i \pi \varepsilon$ governs the dichroic properties of the vacuum and appears once $\eta >1$. In Sec.~\ref{sec:diagrams} we will discuss how the imaginary part appears in a direct evaluation of the Feynman diagrams in Fig.~\ref{fig:diag_three}.

\end{widetext}

\section{Locally-constant field approximation}\label{sec:lcfa}

Here we will employ relatively simple closed-form expressions treating the external background as locally constant. There are basically two different approaches. The first one is based on calculating the polarization tensor in constant crossed fields and the using the actual spatiotemporal dependence of the plane-wave field~\eqref{eq:A_gen} when integrating over $\varphi$. The second method employs the Heisenberg-Euler effective Lagrangian computed in a constant electromagnetic field and takes into account the leading-order quantum correction with respect to the field amplitude $E_0$. The first approach is generally more accurate as it incorporates the higher-order terms in $E_0$ and involves the expression for the polarization operator which is derived for arbitrary photon energies $q_0$. The second technique based on the Heisenberg-Euler Lagrangian is only valid for sufficiently low photon energies, when there is only a small momentum transfer into the $e^+e^-$ loop in the diagram in Fig.~\ref{fig:diag_gen}. Besides, the applicability of this method is limited since it involves the PT expansion with respect to the field amplitude. In what follows, we will describe the both approaches and then thoroughly analyze their validity.

\subsection{Polarization operator in constant crossed fields} \label{sec:lcfa_po}

In the setup under consideration, the vector potential~\eqref{eq:A_gen} is assumed to be a monochromatic plane wave~\eqref{eq:A_pot}. If one replaces $\sin~\varphi$ in Eq.~\eqref{eq:A_pot} with $\varphi$, the external background will obviously become a combination of {\it constant crossed} electric and magnetic fields, $E_x = B_y = -E_0$. In this case, one can also perform nonperturbative calculations of the polarization tensor~\cite{narozhnyi_28_371_1969, batalin_shabad_1968, ritus_ann_phys_1972} and then locally approximate a generic external background by constant crossed fields~\cite{meuren_2013}. Applying this technique to the field configuration~\eqref{eq:A_pot}, one obtains
\begin{eqnarray}
M^{11}_\text{LCFA} &=& \frac{1}{3\pi \xi^2} \int \limits_{-1}^{1} dv \, \bigg ( \frac{\chi}{w} \bigg )^{2/3} \big ( w-1 \big ) g(v),  \label{eq:M11_LCFA} \\
M^{22}_\text{LCFA} &=& \frac{1}{3\pi \xi^2} \int \limits_{-1}^{1} dv \, \bigg ( \frac{\chi}{w} \bigg )^{2/3} \big ( w+2 \big ) g(v),  \label{eq:M22_LCFA}
\end{eqnarray}
where $\chi = 2\xi \eta$, $w = 4/(1-v^2)$, and
\begin{eqnarray}
g(v) &=& \int \limits_{-\pi}^{\pi} d \varphi f' (u) (\cos \varphi)^{2/3}, \label{eq:g}\\
u &=& \bigg (\frac{w}{\chi \cos \varphi} \bigg )^{2/3}, \label{eq:u}\\
f(u) &=& i \int \limits_{0}^{\infty} d \tau \mathrm{e}^{-i(u\tau + \tau^3/3)} = \pi \mathrm{Gi}(u) + i\pi \mathrm{Ai}(u). \label{eq:ritus_f}
\end{eqnarray}
Here $\mathrm{Gi}$ and $\mathrm{Ai}$ are the Scorer and Airy functions, respectively.

Note that the integrals in Eqs.~\eqref{eq:M11_LCFA} and \eqref{eq:M22_LCFA} depend only on $\chi$, i.e. the product $\xi \eta$,  which simplifies the further analysis. This fact is a well-known property of the LCFA ~\cite{ritus_1985}. This approximation is well justified if the parameter $\xi$ is sufficiently large, so one can expect that the predictions~\eqref{eq:M11_LCFA} and \eqref{eq:M22_LCFA} significantly differ from the exact nonperturbative result given in Eq.~\eqref{eq:np_elastic} once $\xi \lesssim 1$. This issue will be discussed in detail in Sec.~\ref{sec:results}.

Finally, we present the asymptotic forms of Eqs.~\eqref{eq:M11_LCFA} and \eqref{eq:M22_LCFA} in the case $\chi \ll 1$. One obtains
\begin{eqnarray}
\mathrm{Re} \, M^{11}_\text{LCFA} &=& -\frac{4\chi^2}{45\xi^2} \bigg [ 1 + \frac{1}{4} \, \chi^2 + \mathcal{O} (\chi^4) \bigg ], \label{eq:M11_LCFA_asym_re}\\
\mathrm{Re} \, M^{22}_\text{LCFA} &=& -\frac{7\chi^2}{45\xi^2} \bigg [ 1 + \frac{13}{49} \, \chi^2 + \mathcal{O} (\chi^4) \bigg ], \label{eq:M22_LCFA_asym_re}\\
\mathrm{Im} \, M^{11}_\text{LCFA} &=& -\frac{3\chi^{3/2}}{8\xi^2} \sqrt{\frac{\pi}{2}} \, \mathrm{e}^{-8/(3\chi)} \big [ 1 + \mathcal{O} (\chi) \big ], \label{eq:M11_LCFA_asym_im}\\
\mathrm{Im} \, M^{22}_\text{LCFA} &=& -\frac{3\chi^{3/2}}{4\xi^2} \sqrt{\frac{\pi}{2}} \, \mathrm{e}^{-8/(3\chi)} \big [ 1 + \mathcal{O} (\chi) \big ]. \label{eq:M22_LCFA_asym_im}
\end{eqnarray}
For small $\chi$ the imaginary part is exponentially suppressed corresponding to tiny probabilities of the photon decay. Note that the ratio $\chi/\xi$ coincides with $\varepsilon = 2\eta$ in Eqs.~\eqref{eq:M11_asym_low_en} and \eqref{eq:M22_asym_low_en}, so the leading-order contribution is reproduced by the LCFA. Nevertheless, the validity of the LCFA and that of the PT expansion correspond to substantially different domains of parameters. Whereas for given $\xi$ they both are accurate for sufficiently small $\eta < \eta_\text{max} (\xi)$, with increasing $\xi$ the bound $\eta_\text{max} (\xi)$ increases in the case of the LCFA and decreases in the case of PT. This will be quantitatively demonstrated in Sec.~\ref{sec:results}. Finally, we note that both the LCFA and PT capture the imaginary part of the polarization tensor.

\begin{widetext}

\subsection{Heisenberg-Euler approximation} \label{sec:lcfa_he}

Another approach is based on the PT expansion of the polarization operator derived from the one-loop effective Lagrangian in the presence of a constant electromagnetic background~\cite{felix_rashid_2015}. The approximate formula for the $\xi^2$ contribution to the polarization tensor has the following form:
\begin{equation}
\Pi^{\mu\nu}_\text{LCFA-HE} (q,k) = \frac{\alpha}{45\pi} \frac{e^2}{m^4} \int \! d^4x \, \mathrm{e}^{i(k-q)x} \Big [ 4 (qF)^\mu (kF)^\nu + 7 (q G)^\mu (k G)^\nu \Big ].
\label{eq:LCFA_gen}
\end{equation}
Here $(kF)^\mu \equiv k_\rho F^{\rho\mu}$. The electromagnetic tensor $F_{\mu \nu} = \partial_\mu \mathcal{A}_\nu - \partial_\nu \mathcal{A}_\mu$ and the dual tensor $G^{\mu \nu} = (1/2) \varepsilon^{\mu\nu\rho\sigma}F_{\rho \sigma}$ are evaluated at the spacetime point $x$ according to the local treatment of the external field. In the case of the plane-wave background~\eqref{eq:A_gen}, the integrals in Eq.~\eqref{eq:LCFA_gen} lead to the conservation laws which may change the photon momentum by $\pm 2\omega$ or keep it the same. We are interested in the latter contribution governing the elastic process. The explicit form of Eq.~\eqref{eq:LCFA_gen} then reads
\begin{equation}
\Pi_\text{LCFA-HE, elastic}^{\mu \nu} (q, k) = \frac{32\pi^3\alpha}{45} m^2 \xi^2 \bigg ( \frac{\omega q_0}{m^2} \bigg )^{2} \delta (k-q)  \begin{pmatrix} 0 & 0 & 0 & 0 \\ 0 & 4 & 0 & 0 \\ 0 & 0 & 7 & 0 \\ 0 & 0 & 0 & 0 \end{pmatrix}.
\label{eq:lcfa_elastic} 
\end{equation}
This exactly corresponds to the leading low-energy terms in Eqs.~\eqref{eq:M11_asym_low_en} and \eqref{eq:M22_asym_low_en} and to the leading-order terms in Eqs.~\eqref{eq:M11_LCFA_asym_re} and \eqref{eq:M22_LCFA_asym_re}. In what follows, they will be denoted by $M^{11}_\text{LCFA-LO}$ and $M^{22}_\text{LCFA-LO}$, respectively. Note that the leading-order LCFA expressions completely disregard the imaginary part of the polarization tensor, i.e., fail to describe the process of dichroism.

\end{widetext}

\begin{figure}[t]
  \center{\includegraphics[width=1\linewidth]{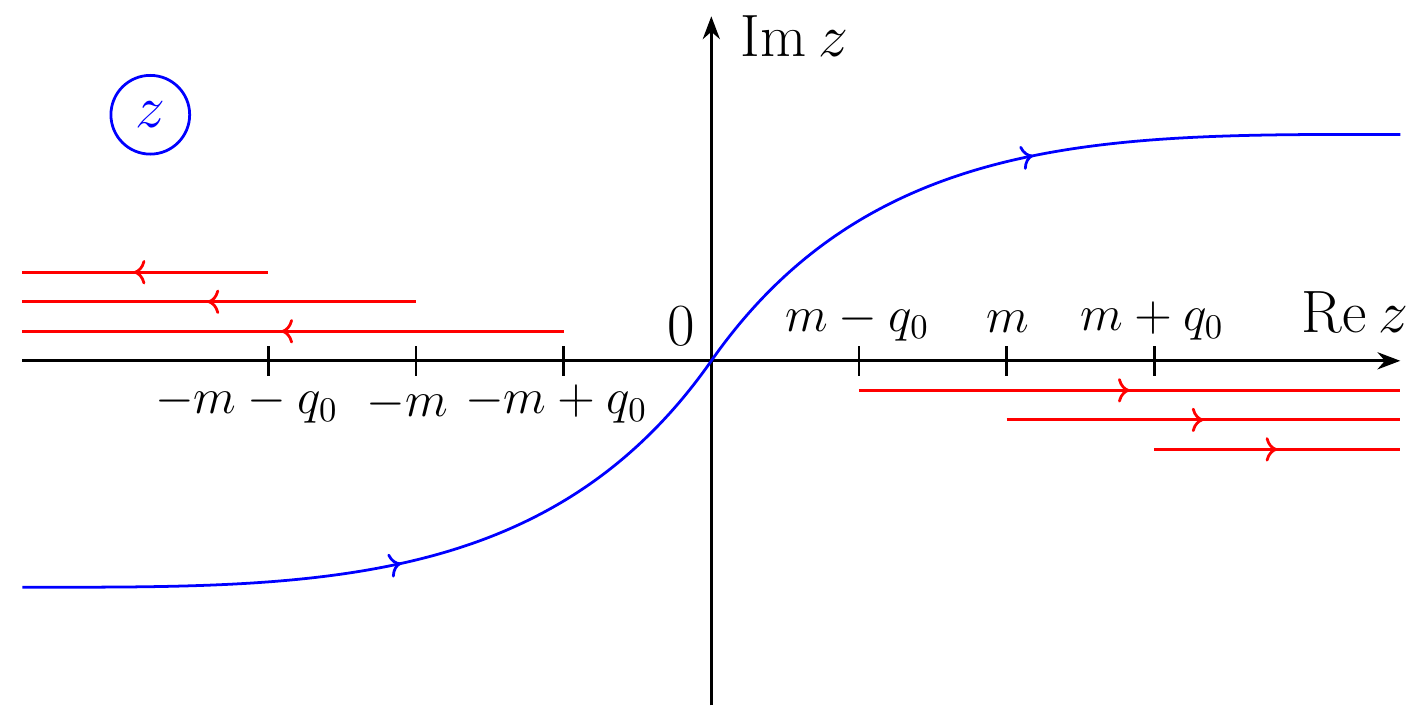}}
  \caption{Branch cuts (red) of the electron propagators in the case $q_0 <m$ before the $z$ integration in Eq.~\eqref{eq:trace_q0} and a possible integration contour (blue).}
  \label{fig:cuts}
\end{figure}

\begin{figure*}[t]
  \center{\includegraphics[height=0.37\linewidth]{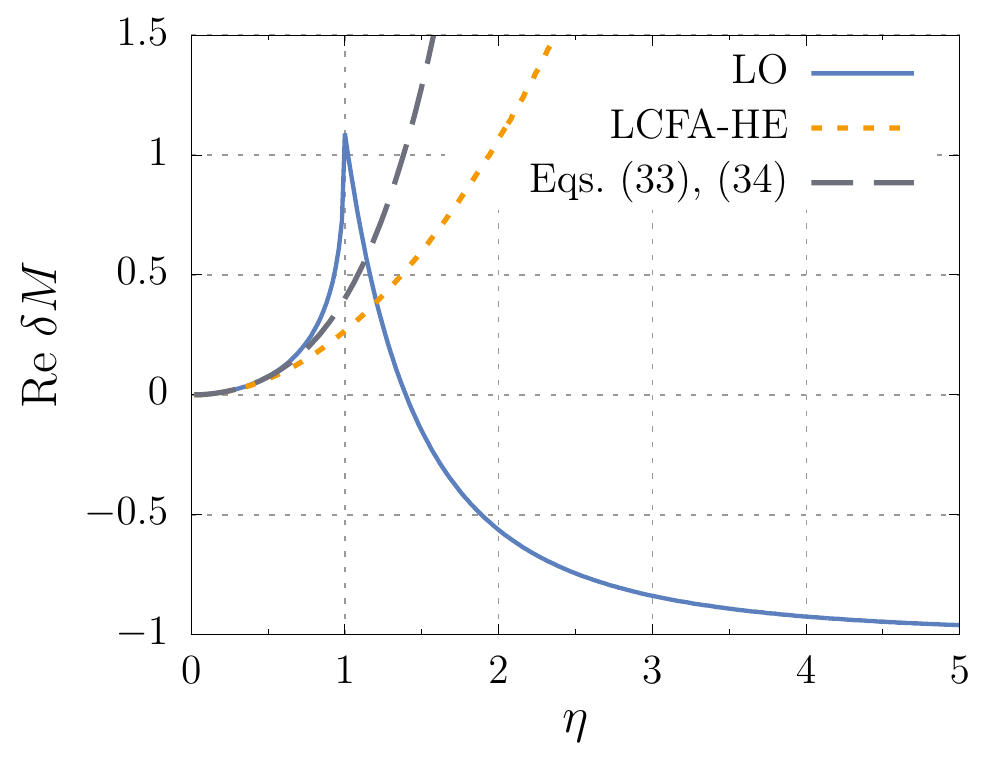}~~~\includegraphics[height=0.37\linewidth]{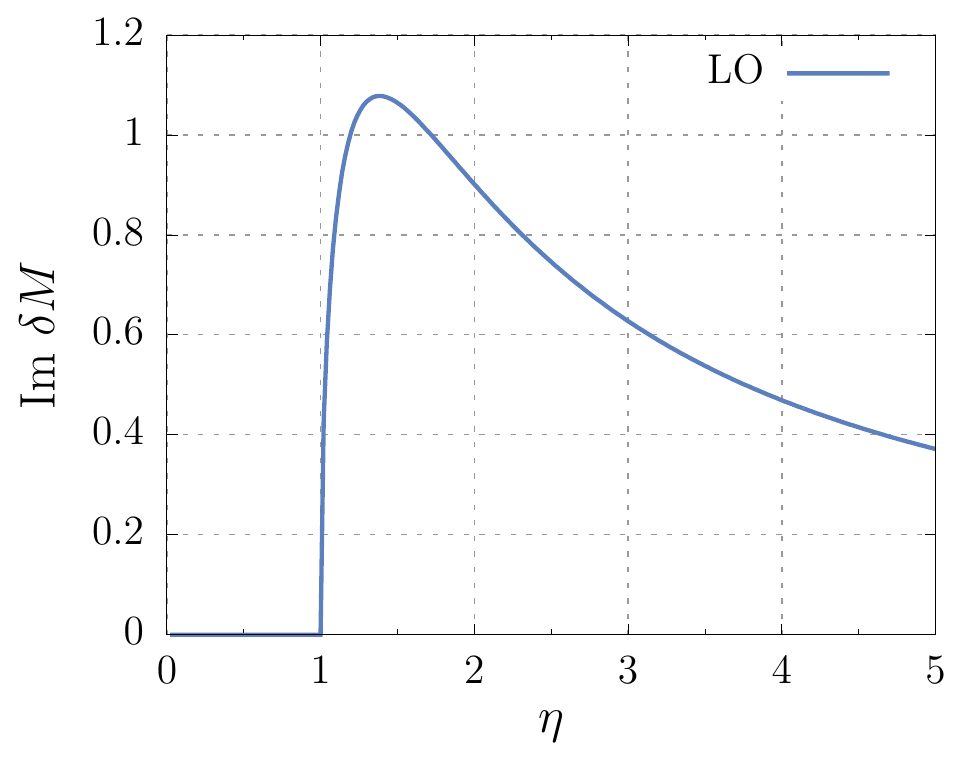}}
  \caption{Real and imaginary parts of the difference $\delta M \equiv M^{11} - M^{22}$ calculated within the leading-order of perturbation theory [Eqs.~\eqref{eq:M11} and \eqref{eq:M22}], by means of the Heisenberg-Euler approximation~\eqref{eq:lcfa_elastic} and according to the low-energy expansions \eqref{eq:M11_asym_low_en} and \eqref{eq:M22_asym_low_en}. The latter two approaches yield zero imaginary part.}
  \label{fig:LO}
\end{figure*}

\section{Direct evaluation of the Feynman diagrams}\label{sec:diagrams}

Here we will directly compute the Feynman diagrams depicted in Fig.~\ref{fig:diag_three}. The corresponding amplitudes and accordingly the contributions to the polarization tensor are proportional to $E_0^2$, i.e. $\xi^2$ [cf.~Eq.~\eqref{eq:M_gen}]. Each interaction vertex involves the energy-momentum transfer with the four-vector $\pm K$, where $K^\mu \equiv \omega n^\mu$ is the four-momentum of the photons that constitute the external plane wave. As we are interested in studying the elastic contributions, the two vertices in each diagram should correspond to one emission and one absorption, so the diagram represents essentially a two-to-two scattering process. Since one has to evaluate three diagrams, the leading-order matrix $M_\text{LO}^{\mu \nu}$ is a sum of three terms, $M_\text{LO}^{\mu \nu} = M_1^{\mu \nu} + M_2^{\mu \nu} + M_3^{\mu \nu}$. Considering, for instance, the first diagram and using the Feynman rules, we obtain the following expression for $M_1^{\mu \nu}$:
\begin{widetext}
\begin{equation}
M_1^{\mu \nu} = -\frac{i}{8\pi^2} \sum_{s=\pm 1} \int d^4p \mathrm{Tr} \, \big [ \gamma^\nu S(p+q/2-sK/2) \gamma^1 S(p+q/2+sK/2) \gamma^\mu S(p - q/2 + sK/2) \gamma^1 S(p-q/2 - sK/2) \big ].
\label{eq:M1}
\end{equation}
\end{widetext}
Here $s$ indicates at which of the two vertices the external-field photon is emitted (absorbed). The integration variables $p^\mu$ are shifted, so that the integrand has a more symmetric form (cf.~Ref.~\cite{OS}). The electron propagator is given by
\begin{equation}
S (p) = \frac{\gamma^\mu p_\mu + m}{m^2 - p^2 - i\varepsilon},
\label{eq:propagator}
\end{equation}
where $\varepsilon \to 0$.

One can explicitly verify that the total expression for $M_\text{LO}^{\mu \nu}$ depends only on the product $\omega q_0$, i.e. $\eta = \omega q_0 /m^2$, in accordance with Eqs.~\eqref{eq:M11} and \eqref{eq:M22}. Therefore, we will assume that $q_0 = \omega = \sqrt{\eta} m$, so $\boldsymbol{K} = -\boldsymbol{q}$. Then Eq.~\eqref{eq:M1} takes the form
\begin{widetext}
\begin{eqnarray}
M_1^{\mu \nu} &=& -\frac{i}{8\pi^2} \int \limits_{-\infty}^{\infty} dz \int d^3\boldsymbol{p} \, \mathrm{Tr} \, \big [ \gamma^\nu S(z, \boldsymbol{p}+ \boldsymbol{q}) \gamma^1 S(z+q_0, \boldsymbol{p}) \gamma^\mu S(z, \boldsymbol{p} - \boldsymbol{q}) \gamma^1 S(z-q_0, \boldsymbol{p}) \nonumber \\
{} &+& \gamma^\nu S(z + q_0, \boldsymbol{p}) \gamma^1 S(z, \boldsymbol{p} + \boldsymbol{q}) \gamma^\mu S(z - q_0, \boldsymbol{p}) \gamma^1 S(z, \boldsymbol{p} - \boldsymbol{q}) \big ].
\label{eq:trace_q0}
\end{eqnarray}
\end{widetext}
The trace contains denominators that for each $\boldsymbol{p}$ turn to zero at complex points $z$ with small nonzero imaginary parts for nonzero values of $\varepsilon$. After the $\boldsymbol{p}$ integration, the trace as a function of $z$ possesses six branch cuts depicted in Fig.~\ref{fig:cuts} for $q_0 <m$. The $z$ integration over the real axis in Eq.~\eqref{eq:trace_q0} can be, in fact, performed over any contour like that displayed in Fig.~\ref{fig:cuts}, provided it does not intersect any of the branch cuts. In the case $q_0 < m$ ($\eta <1$), one can, for instance, rotate the contour, so that it coincides with the imaginary axis. Substituting then $z = i w$, where $w \in \mathbb{R}$, one can explicitly demonstrate that the total contribution $M_\text{LO}^{\mu \nu} = M_1^{\mu \nu} + M_2^{\mu \nu} + M_3^{\mu \nu}$ is real in accordance with Eqs.~\eqref{eq:M11} and \eqref{eq:M22}.

In order to address the high-energy case $\eta > 1$, we employ the following numerical procedure. We change the order of the $z$ and $\boldsymbol{p}$ integrations and first integrate over $z \in \mathbb{R}$. Accordingly, the $z$ integrand has a number of isolated poles $\xi_j - i \sigma \varepsilon$ where $\sigma = \pm 1$ and the real parts $\xi_j$ depend on $\boldsymbol{p}$. In each vicinity $(\xi_j - \delta, \xi_j + \delta)$ we perform the integration semi-analytically by means of the Sokhotski–Plemelj identity. This allows us to set $\varepsilon = 0$ while performing the rest integrations numerically and avoid computational singularities.

Our procedure was also generalized to compute the diagrams for arbitrary independent $q_0$ and $\omega$. The main steps here are generally the same. After that, we confirmed the results obtained by means of the technique described above. Finally, we note that the expression~\eqref{eq:trace_q0} has a similar form to the amplitude of photon emission via the so-called tadpole diagram (see Ref.~\cite{aleksandrov_prd_2019_2}, where it was evaluated in the regime $\eta<1$).

\begin{figure*}[p]
  \center{\includegraphics[height=0.37\linewidth]{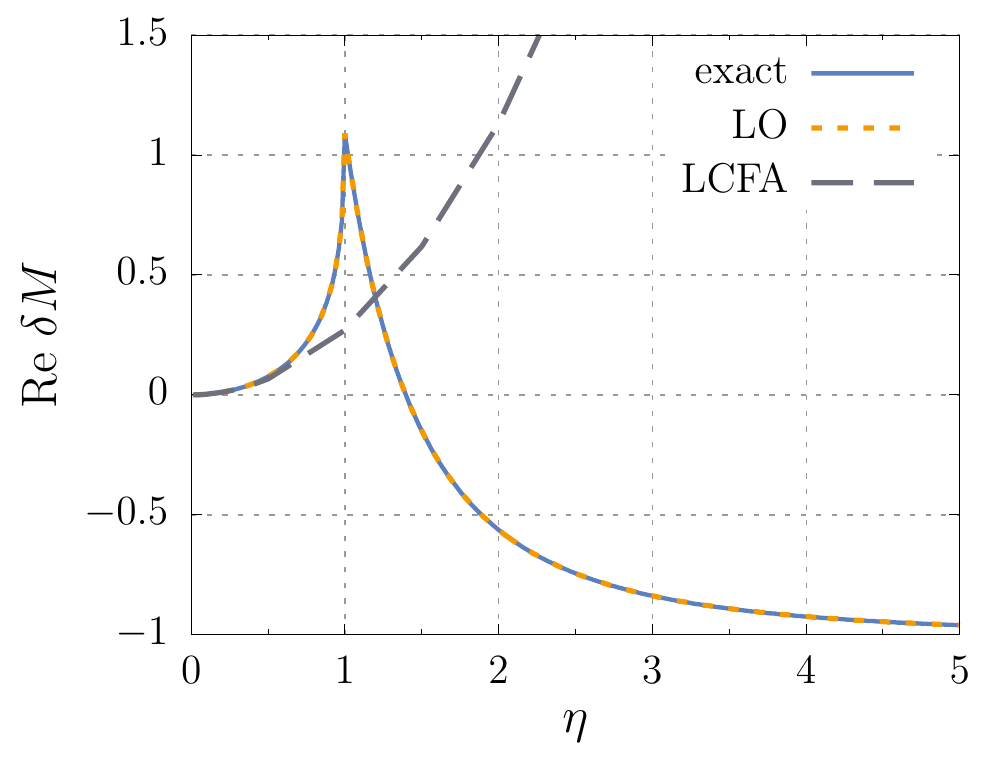}~~~\includegraphics[height=0.37\linewidth]{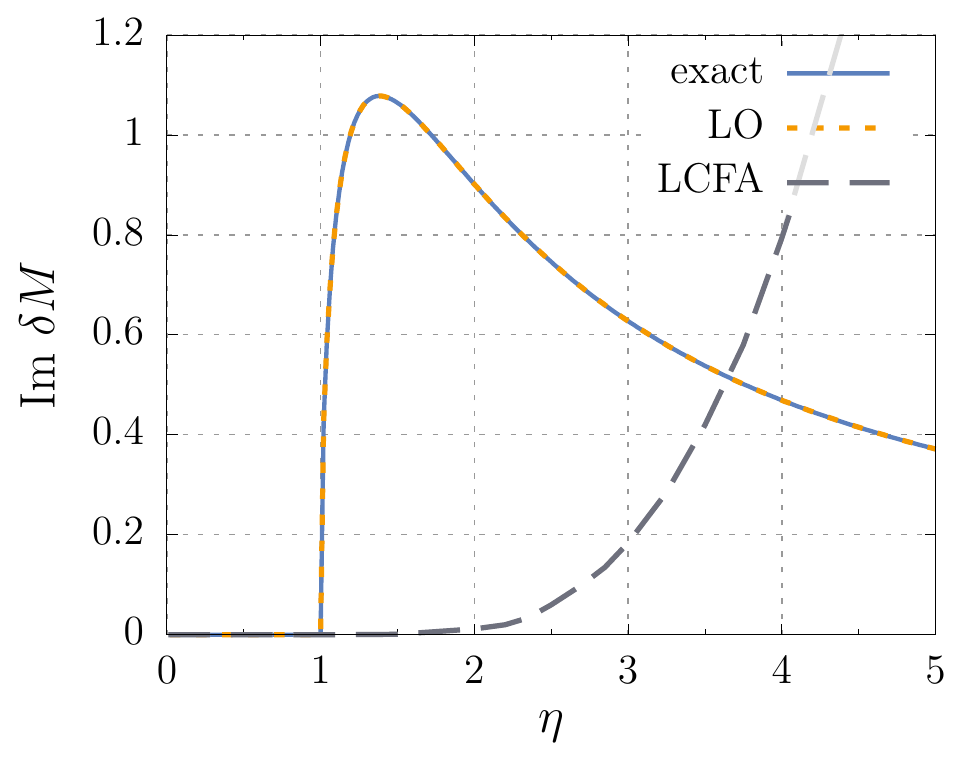}}
  \center{\includegraphics[height=0.37\linewidth]{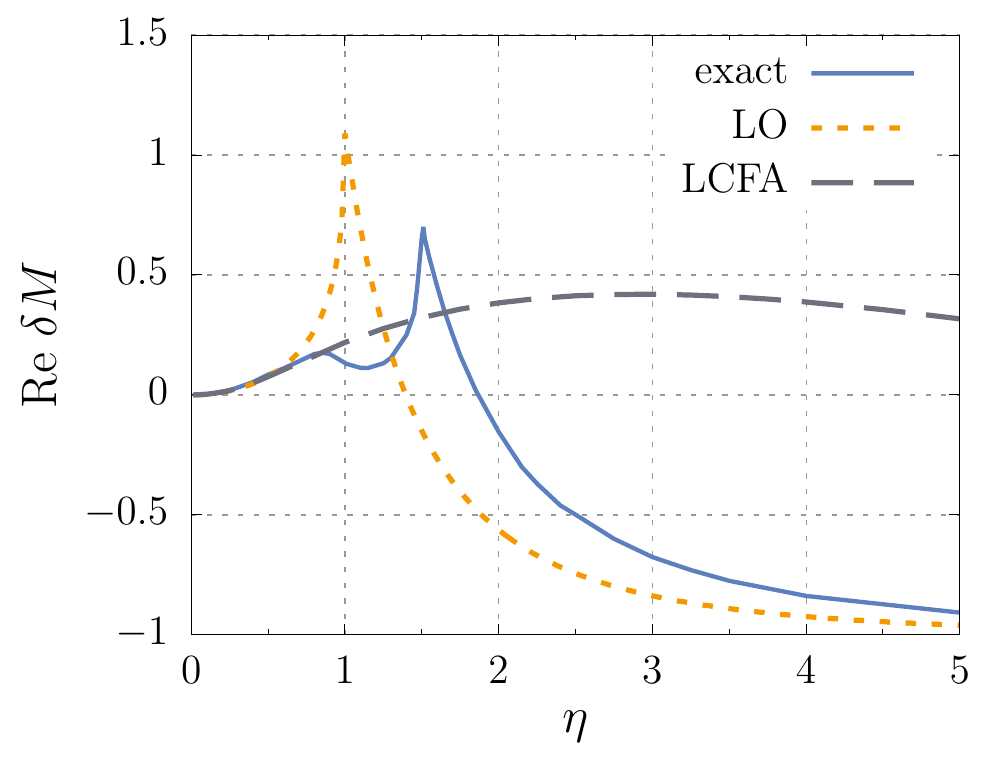}~~~\includegraphics[height=0.37\linewidth]{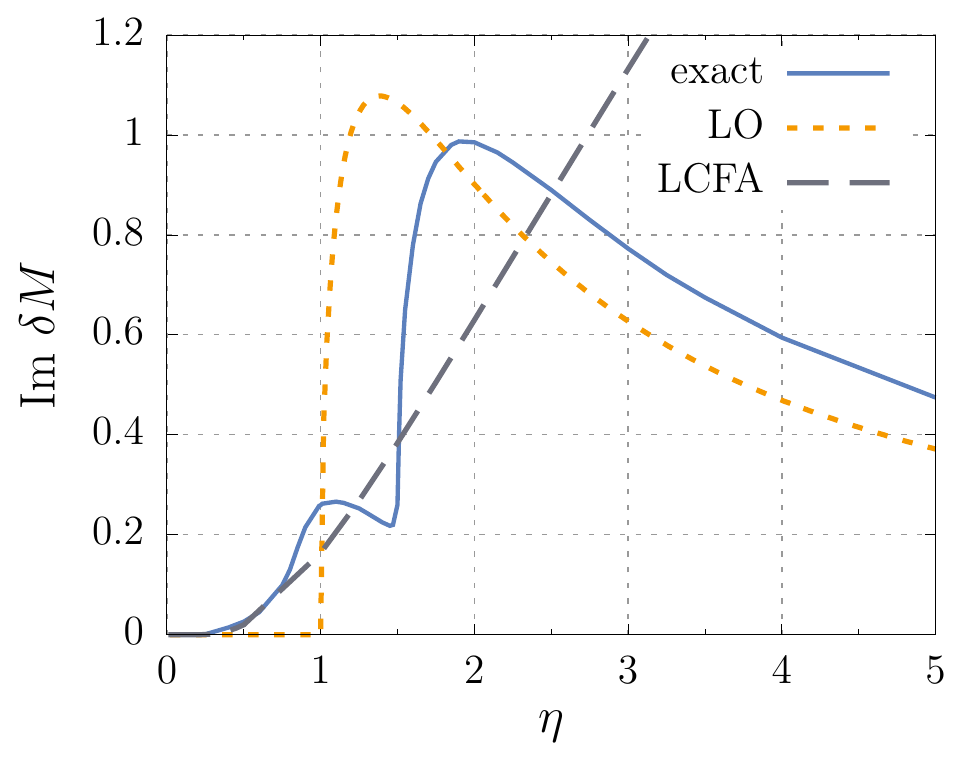}}
  \center{\includegraphics[height=0.37\linewidth]{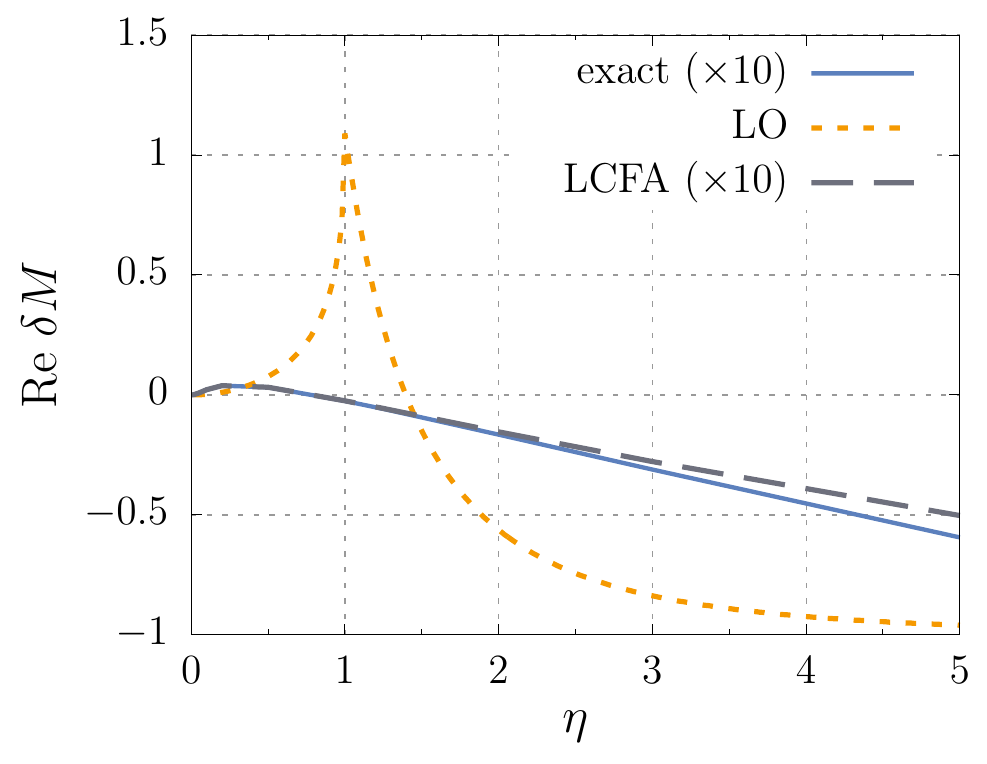}~~~\includegraphics[height=0.37\linewidth]{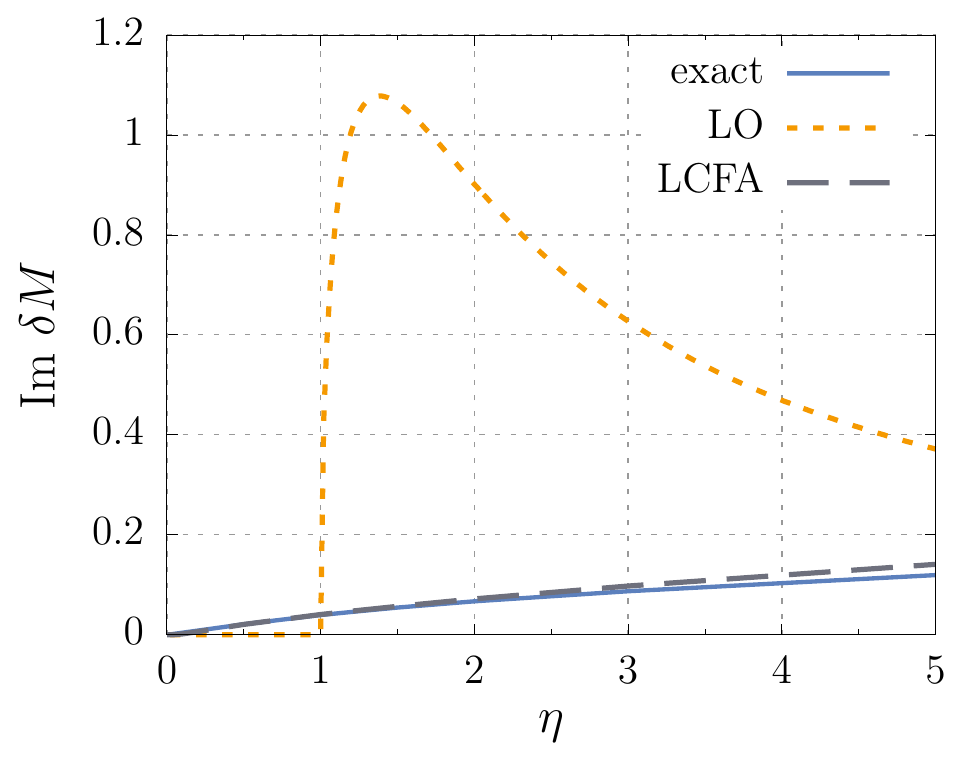}}
  \caption{Real and imaginary parts of the difference $\delta M \equiv M^{11} - M^{22}$ evaluated within the leading-order of perturbation theory (LO), by means of the LCFA [Eqs.~\eqref{eq:M11_LCFA} and \eqref{eq:M22_LCFA}] and exact nonperturbative expression~\eqref{eq:np_elastic} for $\xi = 0.1$ (top), $\xi = 1.0$ (middle), $\xi = 10.0$ (bottom). For $\xi=0.1$ the ``LO'' and exact curves coincide.}
  \label{fig:xi01}
\end{figure*}

\section{Numerical results}\label{sec:results}

We will now perform numerical calculations of the difference $\delta M \equiv M^{11} - M^{22}$, whose real and imaginary parts govern the effects of vacuum birefringence and dichroism, respectively. First, we will evaluate $\delta M$ within the leading order with respect to the field amplitude. In this case, the results do not depend on $\xi$. In Fig.~\ref{fig:LO} we present $\delta M$ as a function of $\eta$. First, one observes that the Heisenberg-Euler approximation within the leading order of perturbation theory can be accurate only in the low-energy regime. If one takes into account the $1/\eta^4$ terms according to Eqs.~\eqref{eq:M11_asym_low_en} and \eqref{eq:M22_asym_low_en}, the results become slightly more accurate although they completely fail to reproduce the full PT results for $\eta>1$. Second, the more general expressions~\eqref{eq:M11} and \eqref{eq:M22} yield a nonzero imaginary part for $\eta >1$, so the PT approach may allow one to describe the effects of dichroism. Finally, we note that our approach based on direct computations of the Feynman diagrams as described in Sec.~\ref{sec:diagrams} provides exactly the same results as Eqs.~\eqref{eq:M11} and \eqref{eq:M22}, which benchmarks the corresponding numerical procedures. To judge whether the leading-order approximation is justified, one has to perform nonperturbative calculations for various values of $\xi$, which will be done next.

\begin{figure}[t]
  \center{\includegraphics[width=0.98\linewidth]{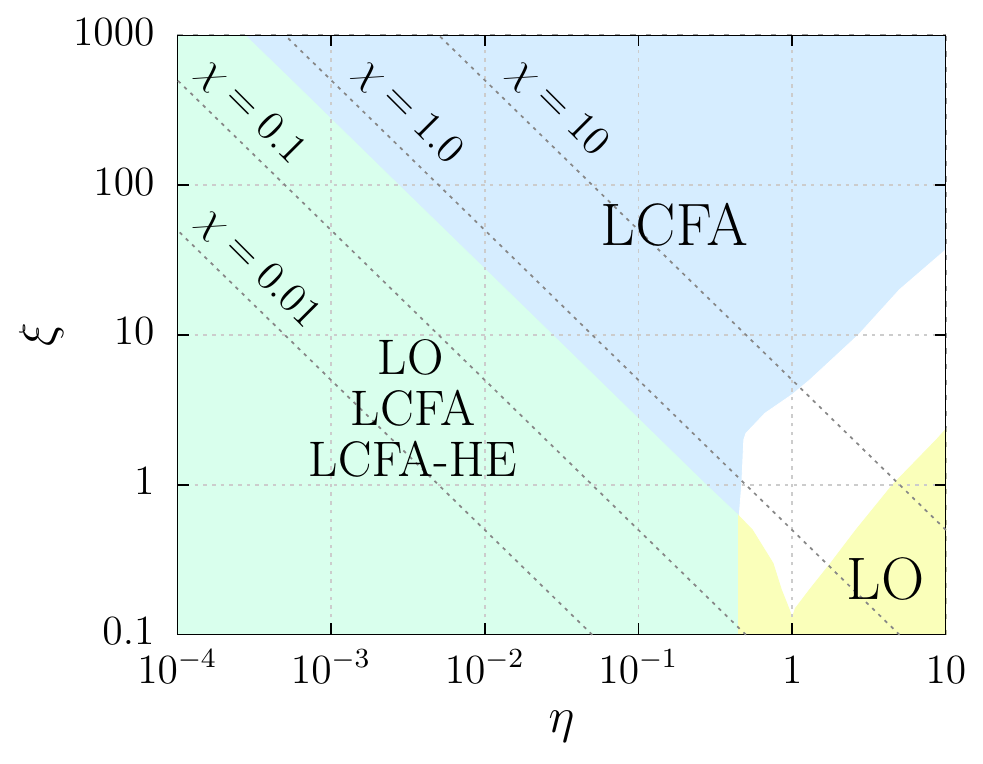}}
  \caption{Domains of the validity of the various approximate methods: LCFA based on using the polarization tensor in constant crossed fields (LCFA), Heisenberg-Euler approximation (LCFA-HE), and PT calculations within the leading order in terms of the external-field amplitude (LO).}
  \label{fig:domains}
\end{figure}

In Fig.~\ref{fig:xi01} we display the real and imaginary parts of $\delta M$ as a function of $\eta$ for three different values of $\xi$: 0.1, 1.0, and 10.0. We refer to Eq.~\eqref{eq:np_elastic} as the {\it exact} result. First, we observe that the $\eta$ dependence very nontrivially changes as a function of $\xi$, which cannot be taken into account by means of the PT approach. Whereas for $\xi \ll 1$, this approximation provides indeed very accurate results within a broad range of $\eta$, for $\xi \gtrsim 1$, it fails to reproduce the exact values unless $\eta \ll 1$. Second, as was mentioned above, the LCFA predictions have the form $\delta M_\text{LCFA} (\xi, \eta) = (1/\xi^2) \delta M_\text{LCFA} (1, \xi \eta)$, so the different LCFA curves can be obtained by simply rescaling the plot axes. This approach does not allow one to describe the nontrivial structure that takes place for $\xi \lesssim 1$, although it is accurate for very small $\eta$, where the expansions~\eqref{eq:M11_LCFA_asym_re} and \eqref{eq:M22_LCFA_asym_re} are valid.

Let us now quantitatively identify the domains of validity of various approximations for describing the vacuum birefringence effects. In Fig.~\ref{fig:domains} we identify the values of $\xi$ and $\eta$ for which the approximate predictions match the exact results with a relative uncertainty on the level of $10\%$. First, let us discuss the PT approach, which yields the leading-order estimates~\eqref{eq:M11} and \eqref{eq:M22}. In the regime $\xi \gg 1$, it is only valid for $\eta \ll 1$. It turns out that in the corresponding domain of parameters $\chi \lesssim 0.5$. Since for large values of $\xi$ one can employ the LCFA, it is possible to estimate the exact result for the real part of $M^{\mu \nu}$ by means of Eqs.~\eqref{eq:M11_LCFA_asym_re} and \eqref{eq:M22_LCFA_asym_re}. Comparing these with the low-energy asymptotic expansions~\eqref{eq:M11_asym_low_en} and \eqref{eq:M22_asym_low_en}, one can obtain the threshold value of $\chi$. For instance, requiring that the relative uncertainty of PT be less than $10\%$, one obtains $\chi < \sqrt{(7/2)0.1} \approx 0.59$. According to our numerical analysis, this condition, in fact, reads $\chi < 0.55$. In the regime $\xi \lesssim 1$, the validity of the LCFA~\eqref{eq:M11_LCFA}, \eqref{eq:M22_LCFA} is very limited, so one has to directly compare the leading-order PT results with the nonperturbative predictions. In this domain, the applicability of perturbation theory is not solely governed by $\chi$ as can be seen in Fig.~\ref{fig:domains}, where the domain of the PT applicability is no longer bounded by a straight line. Finally, we note that in the region $\xi \lesssim 1$, even if the PT approach fails to reproduce the exact results for $\eta \sim 1$, it may provide quite accurate predictions for sufficiently large values of $\eta$, where $\mathrm{Re}~\delta M$ becomes close to $-1$ [see Fig.~\ref{fig:xi01} (middle)]. Moreover, in this region the nonzero imaginary part of the polarization operator can also be obtained by means of perturbation theory.

In order to identify the validity domain of the leading-order Heisenberg-Euler approximation~\eqref{eq:lcfa_elastic}, it is sufficient to compare its predictions with the leading-order PT result~\eqref{eq:M11}, \eqref{eq:M22}. Since within these approaches the matrix $M^{\mu\nu}$ is independent of $\xi$, one should only determine the threshold value of $\eta$. For the $10\%$ uncertainty level, it amounts to $\eta_\text{max} \approx 0.44$. The validity domain of the Heisenberg-Euler approximation is then the intersection of the region $\eta < 0.44$ and the validity domain of the PT approach.

The applicability of the LCFA~\eqref{eq:M11_LCFA}, \eqref{eq:M22_LCFA} corresponds to a much larger region than that where the Heisenberg-Euler approximation is justified. It not only describes the effect of birefringence in the low-energy domain but is also valid in the case of high-energy probe photons ($\eta \gtrsim 1$), provided $\xi \gg 1$.

As was indicated above, the imaginary part of the polarization tensor, which is responsible for dichroic properties of the vacuum, cannot be estimated by means of the leading-order Heisenberg-Euler approximation~\eqref{eq:lcfa_elastic}. Nevertheless, both the PT approach and the LCFA~\eqref{eq:M11_LCFA}, \eqref{eq:M22_LCFA} are very useful here --- they can be employed within the corresponding regions indicated in Fig.~\ref{fig:domains}.

According to our results, the validity domain of the Heisenberg-Euler approximation is the smallest. The corresponding results can always be additionally confirmed by either perturbation theory or the LCFA based on the calculation of the polarization operator in constant crossed fields. The advantage of the latter approach is the possibility to consider larger values of $\eta$ once $\xi \gtrsim 1$. Note also that a considerable part of the plot in Fig.~\ref{fig:domains} relates to large values of the parameter $\chi$, which are not realistic at present. Nevertheless, given the logarithmic scale in the graph, the LCFA covers a domain of parameters which is substantially broader than the validity region of the Heisenberg-Euler approximation. The PT approach is always accurate once the LCFA-HE technique is justified. In addition, the leading-order predictions coincide with the exact results for any values of $\eta$ if $\xi$ is sufficiently small.

\section{Conclusion} \label{sec:conclusions}

In the present investigation, we examined the effects of vacuum birefringence and dichroism in strong plane-wave backgrounds by means of several theoretical methods allowing one to evaluate the leading one-loop contribution to the polarization operator. First, we employed closed-form expressions exactly incorporating the interaction between the electron-positron field and classical external background depending on the spatiotemporal coordinates. Second, we performed calculations within the leading order with respect to the field amplitude, i.e., by means of perturbation theory. This was done by expanding the nonperturbative result and by means of our numerical method based on a direct evaluation of the leading-order Feynman diagrams. It was found that these two approaches yield identical quantitative predictions both for real and imaginary parts of the polarization tensor. Varying the field parameters and the probe-photon energy, we examined the validity of the perturbative methods. Third, we utilized the locally-constant field approximation (LCFA) in two different forms: Heisenberg-Euler approximation and the technique involving exact expressions for the polarization operator in constant crossed fields. By comparing the approximate predictions with the exact results, we evidently identified the field and probe-photon parameters for which each of the approximate techniques is justified.

An important prospect for future studies is the analogous analysis beyond the plane-wave scenario, where the exact analytical expressions are unknown. In this case, for instance, the applicability of the LCFA may be additionally limited if the external electric and magnetic fields are not crossed in contrast to the field configuration examined in the present investigation.

\begin{acknowledgments}
The study was funded by RFBR and ROSATOM, project No. 20-21-00098. I.A.A. also acknowledges the support from the Foundation for the advancement of theoretical physics and mathematics ``BASIS''.
\end{acknowledgments}


\end{document}